\documentclass[prd, aps, superscriptaddress, twocolumn, preprintnumbers, floatfix, nofootinbib]{revtex4}

\usepackage[dvips]{graphicx}
\usepackage{epsf}
\usepackage{amsmath}
\usepackage{amssymb}

\usepackage{graphicx}% Include figure files
\usepackage{dcolumn}% Align table columns on decimal point
\usepackage{bm}% bold math
\def\be{\begin{equation}}
\def\ee{\end{equation}}
\def\bea{\begin{eqnarray}}
\def\eea{\end{eqnarray}}

\usepackage{color}

\begin{document}
\preprint{CERN-PH-TH/2014-043}
%\date{\today}

\title{Closed String Thermodynamics and a Blue Tensor Spectrum}

\author{Robert H. Brandenberger}
\email{rhb@physics.mcgill.ca}
\affiliation{Department of Physics, McGill University, Montr\'eal, QC, H3A 2T8, Canada}
\author{Ali Nayeri}
\email{nayeri@chapman.edu}
\affiliation{Institute for Quantum Studies and Department of Physics, Chapman University, 
Orange, CA 92866, USA}
\author{Subodh P. Patil}
\email{subodh.patil@cern.ch}
\affiliation{Theory Group, CERN, Case C01600, Geneva, CH-1211, Switzerland}
%\author{Cumrun Vafa}
%\email{vafa@physics.harvard.edu}
%\affiliation{Jefferson Physical Laboratory, Harvard University, Cambridge, MA 02138, USA}

\begin{abstract}
The BICEP-2 team has reported the detection of primordial cosmic microwave background B-mode polarization, with hints of a suppression of power at large angular scales relative to smaller scales. Provided that the B-mode polarization is due to primordial gravitational waves, this might imply a blue tilt of the primordial gravitational wave spectrum. Such a tilt would be incompatible with standard inflationary models, although it was predicted some years ago in the context of a mechanism that thermally generates the primordial perturbations through a Hagedorn phase of string cosmology. The purpose of this note is to encourage greater scrutiny of the data with priors informed by a model that is immediately falsifiable, but which \textit{predicts} features that might be favoured by the data-- namely a blue tensor tilt with an induced and complimentary red tilt to the scalar spectrum, with a naturally large tensor to scalar ratio that relates to both.  
\end{abstract}

\maketitle
\newcommand{\eq}[2]{\begin{equation}\label{#1}{#2}\end{equation}}

\section{Introduction}

The BICEP-2 team just announced the detection of primordial cosmic microwave background (CMB) B-mode polarization, implying a tensor-to-scalar ratio of $r = 0.2 \pm 0.05$ \cite{BICEP}. The positive detection of primordial gravitational waves constitutes a major advance for early universe cosmology, giving us a new diagnostic tool with which to scrutinize models of the very early universe against observational data. Conventional adiabatic cosmological fluctuations do not predict any B-mode polarization at the linear level in cosmological perturbation theory. Hence in the context of the simplest models, primordial B-mode polarization must be due to gravitational waves \footnote{Note, however, that beyond the simplest single field scalar models, there are other sources of B-mode polarization e.g. from cosmic strings \cite{Holder}. B-mode polarization will also be produced by lensing of E-mode polarization, which in turn is directly generated from cosmological fluctuations, and that this B-mode lensing signal has in fact recently been discovered by the South Pole \cite{Hanson} and the Polarbear telescopes \cite{Dobbs}.}. 

\begin{figure}[t]
%\epsfile{file=wakewedge.eps,scale=1.0}
\includegraphics[height=6cm]{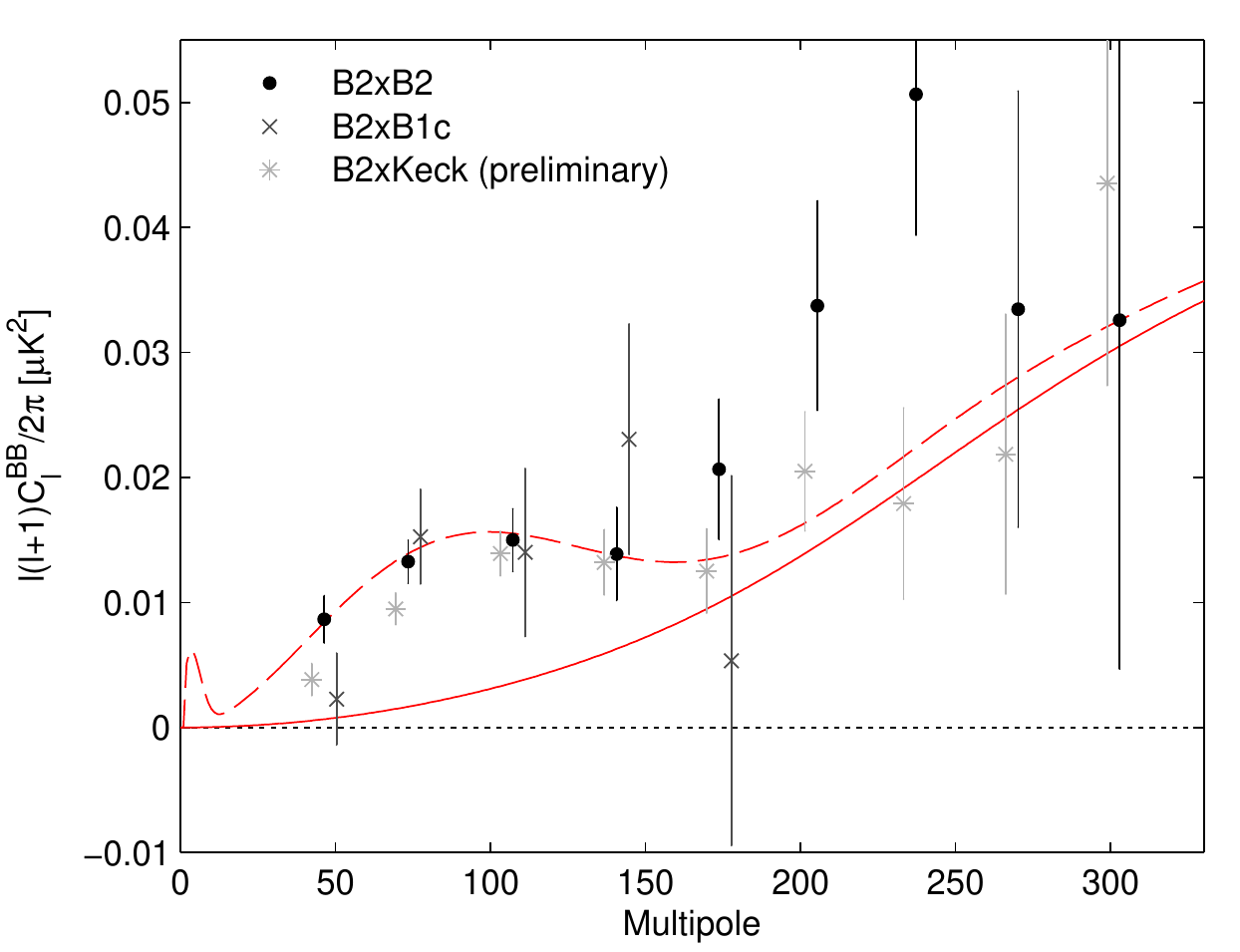}
\caption{The BB auto and cross correlation functions as seen by the BICEP collaboration (courtesy of \cite{BICEP}).} \label{bicepfig}
\end{figure}

Although the BICEP-2 collaboration's analysis took $n_T=0$ as a prior in its simulated data, we wish to ask whether a suppression of power in the BB angular power spectrum at large angular scales relative to smaller scales might be seen in the data, in particular in the B2 x Keck cross correlation function at long wavelengths (which is less sensitive to systematic noise\footnote{Which we note can only boost the auto-correlation function.}) and in the B2 x B2 correlation function, although the latter is more susceptible to contamination from foregrounds (see Fig. \ref{bicepfig}). Whether this suppression is statistically significant remains to be seen. If it is, it could be interpreted as indicative of a positive tilt of the primordial tensor spectrum at the largest angular scales. If this does turn out to be the case, then this result would be very hard to interpret in the context of the standard inflationary paradigm of early universe cosmology (see also \cite{contaldi} for an analysis of the additional tension between measuring a large $r$ with the small $k$ scalar power spectrum).  
%%RB: I slightly toned down the wording in this paragraph.

Assuming that space-time is described by General Relativity and that matter obeys the ``weak energy condition", inflation generically predicts a red spectrum of gravitational waves, i.e. $n_T < 0$. This arises from the fact that the amplitude of the gravitational waves on a scale $k$ is set by the amplitude of the Hubble expansion rate $H$ at the time when that scale exits the Hubble radius, and that during inflation ${\dot{H}} < 0$. For single field slow roll models this relation is precisely 
\be
\label{sftilt}
n_T = -2\epsilon \, ,
\ee
with $\epsilon := -\dot H/H^2$. A related challenge for standard inflationary cosmology in light of the BICEP-2 data is that the tensor-to-scalar ratio $r = 0.2$ implies a large field excursion over the duration in which the observed modes in the CMB were produced \cite{Lyth}:
\be
\Delta\phi \gtrsim M_{pl} \, .
\ee
Constructing a model that safely accomplishes this is challenging to say the least from the perspective of effective field theory, as field excursions comparable to the cut-off of the theory typically generate large anomalous dimensions for operators that were initially suppressed (by appropriate powers of the cutoff), potentially spoiling the requisite conditions for inflation to occur as it progresses\footnote{The so called sensitivity of large field models to `Planck slop' \cite{BCQ}.}. However, this is not to say that this might not be accomplished in the context of some fundamental theory construction--  see \cite{SW} for an interesting claim (and \cite{JC} for a counter-claim)-- within the context of string theory, large field excursions certainly appear to be problematic \cite{swampland}. 
 
In this note we wish to remind cosmologists of a mechanism to generate the primordial perturbations from the thermodynamics of \textit{closed} strings in a quasi-static background, which--
\begin{itemize}
\item Naturally generates a large tensor to scalar ratio;
\item \textit{Predicts} a blue tilt to the tensor spectrum,
\item with a complimentary red tilt to the scalar spectrum, both of which relate to $r$. 
\end{itemize}
This construction relies upon a background that consists of a quasi-static initial state in the Einstein frame, whose specific realization can be addressed in the context of particular string constructions (see \cite{Florakis, KPT1} for some recent attempts), but whose existence we will take for granted in the following as far as the study of fluctuations is concerned, just as one typically does in the context of inflationary cosmology\footnote{Requiring that inflation exists in the context of a consistent quantum theory requires considerable tuning at the level of the low energy effective description \cite{ETA} (its so-called UV sensitivity).}. In fact, this is the very premise of the effective theory of the adiabatic mode \cite{Senatore}-- the so called effective theory of inflation. In what follows, we will first address plausible constructions that could give rise to the requisite background as motivation for the subsequent section-- the main focus of this note-- where we argue that \textit{the thermodynamics of closed strings in the early universe can naturally generate a large, blue titled tensor mode background.} 

Our goal is to provide observations with a novel, predictive, and falsifiable model which can inform the formulation of priors when analyzing the data in a manner that is easily contrasted against the predictions of inflationary cosmology. Whether there are hints for a blue tensor tilt in the data is to be viewed as secondary to the goal of providing a `straw-model' with which to contrast the predictions of inflation against, the scientific utility of which needing no further elaboration. 

\section{Closed string thermodynamics and a quasi static initial universe}

The geometry of string theory is a very rich and complex subject. There exist very distinct geometries, sometimes with very distinct topologies that are indistinguishable from each other as far as physical processes involving strings are concerned. Known as `dualities' \cite{GPR}, the connections between these geometries is one of the most striking features of string theory that persists at low energies, a pervasive manifestation of which is the T-duality symmetry that relates strings in a very large universe (relative to the string scale) to strings in a very small universe. In the absence of any background fluxes, in the context of Heterotic string theory for example, this implies the duality
\be
\label{TD}
G_{ab} \leftrightarrow G^{-1}_{ab}
\ee
Where $G_{ab}$ is the (target space) metric of spacetime. The implications of this duality on early universe cosmology has been studied extensively in various constructions \cite{PBB, TV}. The particular context we are concerned with, ``string gas cosmology", is a paradigm of early universe cosmology initially proposed in \cite{BV} to explain why only three of the nine spatial dimensions of string theory can be macroscopic. Within a particular realization of this framework, given certain assumptions, one can naturally generate a large tensor background with a spectrum that is blue tilted \cite{BNPV}, with a red tilted scalar spectrum \cite{NBV}\footnote{As has been remarked since this model was proposed \cite{SGCrevs}, a detection of a blue spectrum of tensor modes can be viewed as a prediction
for cosmological observations, first made in the context of string theory, that would falsify the inflationary paradigm if obserevd.}. 

The cosmological model we consider \cite{NBV, BNPV} is based on the thermodynamics of \textit{closed} heterotic strings. Due to the existence of an exponential tower of oscillatory string modes, there is a maximal temperature $T_H$ which a thermal gas of strings can attain \cite{Hagedorn}. The existence of winding modes in addition to the center-of-mass momentum modes is the representation of the T-duality (\ref{TD}) on the matter content of the universe, wherein physics on a torus of radius $R$ is equivalent to that on a torus of radius $l_s^2 / R$, where $l_s$ is the string length. This duality leads to the temperature/radius curve for a weakly coupled gas of strings indicated in Figure 2 (where the vertical axis is the temperature, and the horizontal axis the radius $R$ in string units on a logarithmic scale). It thus seems reasonable to conjecture that the cosmological singularity might be dynamically resolved by the energetics of the so called Hagedorn phase\footnote{As has been explicitly demonstrated in the context of type II strings in \cite{KPT1, KPT2}.}.

\begin{figure}[htbp]
%\epsfile{file=wakewedge.eps,scale=1.0}
\includegraphics[height=5cm]{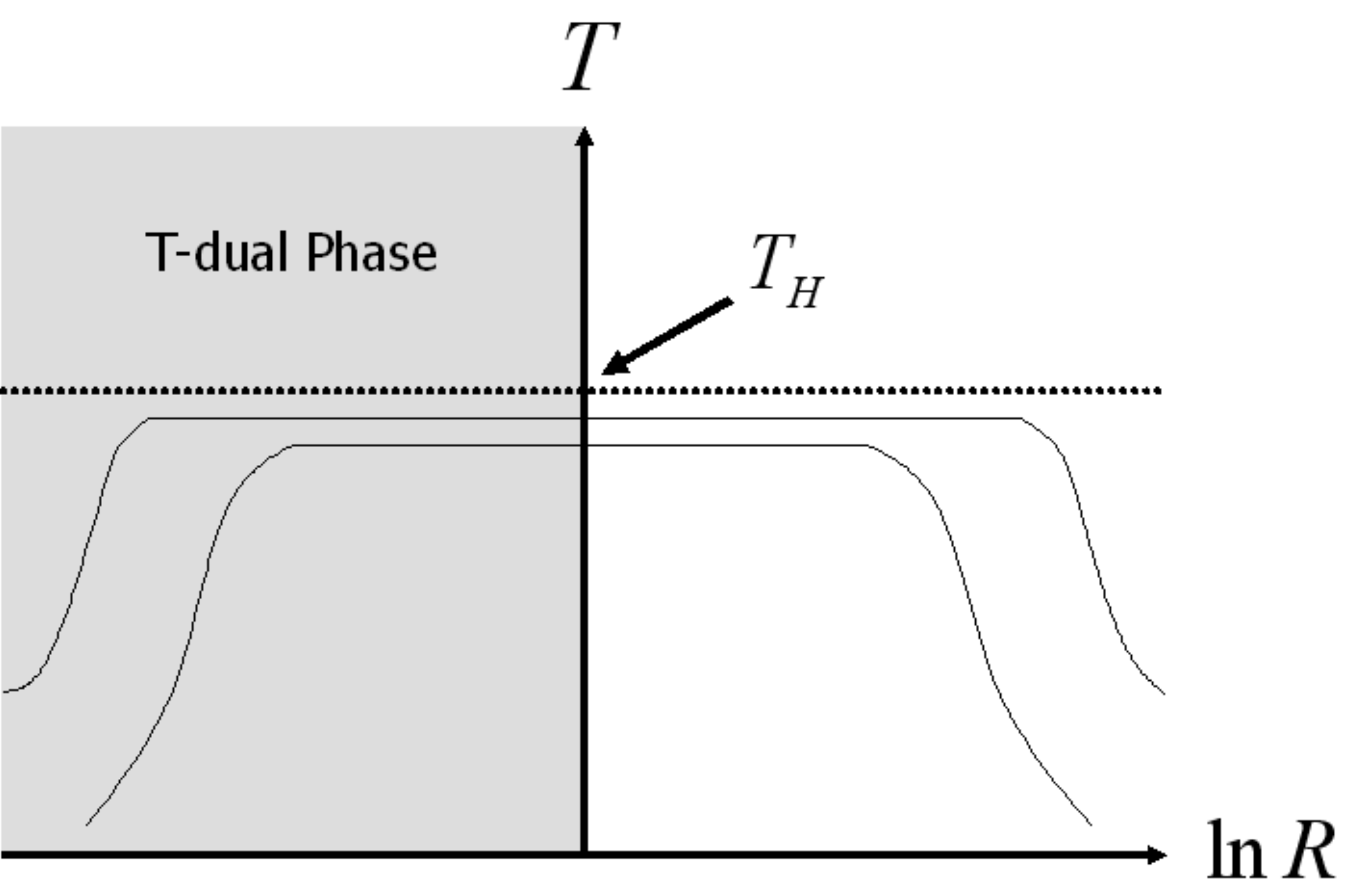}
\caption{Temperature $T$ of a gas of closed strings on a toriodal background 
as a function of the radius $R$ of the torus.} \label{fig:1}
\end{figure}

The model of \cite{NBV, BNPV} is based on the premise that the universe
starts in the quasi-static Hagedorn phase when the temperature
is only very slightly lower than the Hagedorn temperature \footnote{As
explained in \cite{BV}, the temperature difference depends inversely
on the entropy}. The decay of string winding modes will 
eventually enable three spatial dimensions to become large, while
the others are forever confined by string winding modes \cite{BV}
\footnote{The role played by T-duality in ensuring moduli stabilization was discussed in detail in \cite{moduli}.}. The decay of the string
winding modes leads to a smooth transition to the radiation-dominated
phase of Standard Cosmology. The transition time between the quasi-static Hagedorn phase with constant scale factor $a(t)$ and the radiation phase with 
$a(t) \sim t^{1/2}$ is denoted by $t_R$, since it plays a role similar
to the reheating time in inflationary cosmology. 

%\begin{figure}[htbp]
%\epsfile{file=wakewedge.eps,scale=1.0}
%\includegraphics[height=5cm]{timeevol.eps}
%\caption{Time evolution of the scale factor in string gas cosmology.} %\label{fig:2}
%\end{figure}

In Figure 3 we show the evolution of various scales in string gas cosmology.
In this sketch, the vertical axis is time, the horizontal axis represents physical distance. The two light red curves which are vertical in the Hagedorn phase indicate the physical wavelengths of two different fluctuation modes. The solid blue curve which grows linearly in the radiation phase and is at infinity early in the Hagedorn phase is the Hubble radius $l_H(t) = H^{-1}(t)$, the inverse expansion rate $H(t) = {\dot a}/a$ (where the dot indicates the derivative with respect to time $t$). The Hubble radius separates scales on which fluctuations oscillate (sub-Hubble modes) from those where the oscillations are frozen out and the amplitude of the modes is squeezed (see \cite{MFB} for discussions of how cosmological perturbations evolve), namely the super-Hubble modes.

\begin{figure}[htbp]
%\epsfile{file=wakewedge.eps,scale=1.0}
\includegraphics[height=9cm]{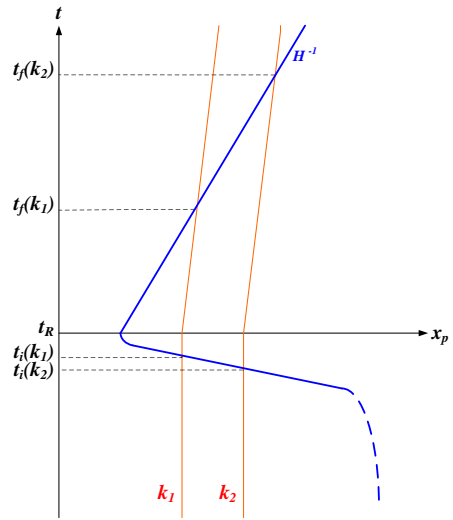}
\caption{Space-time sketch of the evolution in string gas cosmology. The vertical axis is time, the horizontal axis is physical distance. The time $t_R$ corresponds to the transition between the Hagedorn phase and the radiation phase. The thick blue curve labelled by $H^{-1}$ indicates the Hubble radius, the two thin red curves which are vertical during the Hagedorn phase correspond to the physical wavelengths of fluctuation modes labelled by $k_1$ and $k_2$. } \label{fig:3}
\end{figure}
The first point to remark is that the horizon is much larger than the Hubble
radius (in fact infinite if time extends to $- \infty$). Hence, string gas cosmology addresses the horizon problem of standard cosmology in a complimentary way to inflation. Secondly, it is clear that cosmological fluctuations begin on sub-Hubble scales and evolve after $t_R$ for a long time at super-Hubble lengths. The sub-Hubble origin of the scales makes it possible to have a causal generation mechanism of fluctuations, the super-Hubble period of evolution will lead to acoustic oscillations at late times in both the angular power spectrum of CMB anisotropies and in the matter power spectrum \cite{SZ}.

Having set the scene for the background, we now turn to reviewing why this string cosmological background can lead to a spectrum of cosmological perturbations with a red tilt and gravitational waves with a blue tilt, generated by the thermodynamics of strings.  

\section{Tensor Modes from an early Hagedorn phase}

Since in the initial Hagedorn phase of string cosmology matter is a
thermal gas of strings, the initial conditions for scalar and tensor metric
fluctuations are thermal rather than vacuum and the energy-momentum tensor
correlation functions are determined by closed string thermodynamics rather
than by open or point particle thermodynamics \cite{Ali}.

The calculation of the spectrum of scalar \cite{NBV} and tensor \cite{BNPV}
fluctuations in string gas cosmology proceeds in three steps. In the first,
the matter correlation functions are evaluated using the results of the
closed string thermal partition function given in \cite{Deo}. The second
step is to use the Einstein constraint equations, presuming our quasi-static background to be a given \textit{in the Einstein frame}\footnote{From \cite{more} we know this is a non-trivial assumption. However, see \cite{KPT1} for suggestions as to how one could accomplish this in the context of type II superstrings.}, to determine the cosmological fluctuations and gravitational waves from the matter correlation functions mode by mode when the modes $k$ exit the Hubble radius at the times $t_i(k)$. The third step is to evolve the gravitational fluctuations until the present time using the usual theory of cosmological fluctuations. 

The metric including cosmological fluctuations $\Phi({\bf x}, \eta)$ and
gravitational waves $h_{ij}({\bf x}, \eta)$ can be written in the form
\cite{MFB}
\be \label{pertmetric}
d s^2 \, = \, a^2(\eta) \left\{(1 + 2 \Phi)d\eta^2 - [(1 - 
2 \Phi)\delta_{ij} + h_{ij}]d x^i d x^j\right\} \, , 
\ee 
where $\eta$ is conformal time related to physical time via
$dt = a(t) d\eta$. The scalar metric fluctuations are determined
via the energy density perturbations via
\be \label{scalarexp} 
\langle|\Phi(k)|^2\rangle \, = \, 16 \pi^2 G_N^2 
k^{-4} \langle\delta T^0{}_0(k) \delta T^0{}_0(k)\rangle \, , 
\ee 
where the pointed brackets indicate \textit{thermal} expectation values, $T^{\mu}_{\nu}$
is the energy-momentum tensor, and $G_N$ is Newton's gravitational constant. The gravitational waves are given by the off-diagonal (i.e. $i \neq j$) pressure fluctuations: 
\be 
\label{tensorexp} \langle|h(k)|^2\rangle \, = \, 16 \pi^2 G_N^2 
k^{-4} \langle\delta T^i{}_j(k) \delta T^i{}_j(k)\rangle \, .
\ee 

To determine the energy density fluctuations, we use the
fact that in thermal equilibrium the position space
perturbations are given by the
specific heat capacity $C_V$ at fixed volume $V = R^3$
\be \label{cor1b}
\langle \delta\rho^2 \rangle \,  = \,  \frac{T^2}{R^6} C_V \,  .
\ee 
For a thermal gas of heterotic strings $C_V$ is given by
\be \label{specheat2b} 
C_V  \, \approx \, 2 \frac{R^2 / l_s^3}{T \left(1 - T/T_H\right)}\, , 
\ee 
and hence the power spectrum of $\Phi$, defined by
\be \label{power2} 
P_{\Phi}(k) \,  \equiv  \, {1 \over {2 \pi^2}} k^3 |\Phi(k)|^2 \,
\ee
is determined to be \cite{NBV}
\be \label{power3}
P_{\Phi}(k) \, = \,  \left(\frac{l_{Pl}}{l_s}\right)^4{T(k) \over {T_H}} {1 \over {1 - T(k)/T_H}} 
\,  ,
\ee 
where $T(k)$ is the temperature at the time $t_i(k)$ when mode $k$
exits the Hubble radius.

As inferred from Figures 2 \& 3, the temperature $T(k)$ decreases
as $k$ increases, since large $k$ modes exit the Hubble radius later.
Since $T(k)$ is close to the Hagedorn temperature, it is the denominator
of the right hand side of (\ref{power3}) which dominates the final
amplitude. Hence, the spectrum of scalar metric fluctuations has a
red tilt (larger amplitude at larger wavelengths). Neglecting running, the tilt can be computed as (defining $\widehat T(k) := T(k)/T_H$)
\be
\label{scalartilt}
n_s-1 = (1- \widehat T(k))^{-1}k \frac{d \widehat T(k)}{d k},
\ee
which is negative since $d\widehat T/dk < 0$, and arbitrarily small in the limit of a sudden transition (in which case $d\widehat T/dk \equiv 0$). The power spectrum of the tensor modes, which is produced by fluctuations of the wound strings around a compact space, is given by (\ref{tensorexp})
the correlation function $C^i{}_j{}^i{}_j(R)$ ($i \neq j$) , namely the mean 
square fluctuation of $T^i{}_j$ ($i \neq j$) in a region of radius $R = k^{-1}$
\be \label{tpower2}
P_h(k) \, = \, 16 \pi^2 G_N^2 k^{-4} C^i{}_j{}^i{}_j(R) \, .
\ee
The correlation function $C^i{}_j{}^i{}_j$ on the right hand side
of the above equation follows from the thermal closed string
partition function and was computed in \cite{Ali, BNPV2} (see also \cite{new} for a more general treatment), with the result that for temperatures close to the Hagedorn value
\be \label{tresult}
P_h(k) \, \sim \,
\left(\frac{l_{Pl}}{l_s}\right)^4 \frac{T(k)}{T_H}(1 -
T(k)/T_H)\ln^2{\left[\frac{1}{l_s^2 k^2}(1 - T(k)/T_H)\right]} \, .
\ee
The key factor $(1 - T(k)/T_H)$ now appears in the numerator
and hence leads to a blue spectrum. Neglecting running (and thus the logarithmic factor as well), the tilt can be computed as
\bea
n_T &=& \frac{1 - 2\widehat T(k)}{1- \widehat T(k)}k\, \frac{d \widehat T(k)}{d k} \nonumber\\ \label{tfinal}
&=& -(n_s-1)(2\widehat T(k) - 1) \, ,
\eea
where we see the complimentarity between the tilt of the scalar and tensor spectra. The fact that we obtain a blue spectrum of gravitational waves is
readily understood. The spectrum of gravitational waves
is determined by the anisotropic pressure perturbations. Since deeper
in the Hagedorn phase, i.e. at higher $T(k)$, the pressure is smaller,
the anisotropic pressure fluctuations should be smaller, as well. Hence,
the amplitude of the gravitational wave spectrum will increase towards
the ultraviolet, corresponding to a blue spectrum. Furthermore, we can also compute the tensor to scalar ratio as
\be
r = (1 - \widehat T)^2 \ln^2{\left[\frac{1}{l_s^2 k^2}(1 - \widehat T(k))\right]} \, .
\ee
Requiring COBE normalization for the power spectrum for the comoving curvature perturbation \cite{NBV}, in addition to requiring a tensor to scalar ratio of $0.2$ \cite{BICEP} fixes the string length to be given by $l_{Pl} = 0.0016\, l_s$, and that the modes we observe exited when the temperature of the universe was $T \sim 0.99 T_H$. The latter implying that the tensor tilt is essentially equal and opposite to the scalar tilt
\be
n_T \approx -(n_s-1) \, ,
\ee
the precise value of which depends on the manner in which the background exited the Hagedorn phase.  
 
\section{Discussion}

In generating the primordial perturbations from the thermodynamics of closed strings, one can isolate the background dependence to one free parameter, the ratio of string length and Planck length, and a function $T(k)$ which is close to the Hagedorn temperature and is a decreasing function of $k$. The precise $k$ dependence depends of $T$ encodes the details of the transition between the Hagedorn phase and the radiation phase, and will determine the relative tilts of the spectra (\ref{scalartilt}) and (\ref{tfinal}), which are to first approximation, equal and opposite (\ref{tfinal}). With these inputs we can also compute the amplitudes of the scalar and tensor spectra at any pivot scale, which is unique for only \textit{three spatial large compact dimensions} \cite{Ali}.  As emphasized in \cite{BNPV}, the key result is that the tensor spectrum has a blue tilt, whereas the scalar fluctuations retain a red tilt. This feature distinguishes the string thermodynamic generation of the primordial perturbations from standard inflationary realizations. 

%Note that subsequent to our work \cite{BNPV}, other cosmological models were proposed which yield a blue tilt of gravitational waves. One possibility is to invoke super-inflation (a period with ${\dot H} > 0$). However, new physics violating the weak energy condition is required in these models \cite{Biswas} \cite{LQC}.

\acknowledgements{We wish to thank Matt Dobbs, Gil Holder and Cumrun Vafa for valuable discussions and correspondence. SP is supported by a Marie Curie Intra-European Fellowship of the European Community's 7'th Framework Program under contract number PIEF-GA-2011-302817. RB is supported by an NSERC Discovery Grant, and by funds from the Canada Research Chair program.}

\end{document}